\documentstyle[12pt]{article}
\def\Tr{{\rm Tr}}
\def\a{\begin{eqnarray}}
\def\b{\end{eqnarray}}
\def\0{\nonumber}
\def\ba{\begin{array}}
\def\ea{\end{array}}

\def\ed{{e^{\partial}}}
\def\edk{{e^{k\partial}}}
\def\edkd{{e^{(k-2)\partial}}}
\def\edkt{{e^{(k-3)\partial}}}
\def\edkp{{e^{(k-4)\partial}}}
\def\edkc{{e^{(k-5)\partial}}}
\def\em{e^{-\partial}}
\def\emd{e^{-2\partial}}
\def\emt{e^{-3\partial}}
\def\emp{e^{-4\partial}}

\def\ak{\hat A^{[k]}}
\def\an{\hat A^{[N]}}
\def\bak{\bar A^{[k]}}
\renewcommand{\theequation}{\thesection.\arabic{equation}}

\newlength{\extraspace}
\newlength{\extraspaces}
\newcounter{dummy}
\newcommand{\ai}{
\addtocounter{equation}{1}
\setcounter{dummy}{\value{equation}}
\setcounter{equation}{0}
\renewcommand{\theequation}{\thesection.\arabic{dummy}\alph{equation}}
\begin{eqnarray}
\addtolength{\abovedisplayskip}{\extraspaces}
\addtolength{\belowdisplayskip}{\extraspaces}
\addtolength{\abovedisplayshortskip}{\extraspace}
\addtolength{\belowdisplayshortskip}{\extraspace}}
\newcommand{\bj}{
\end{eqnarray}
\setcounter{equation}{\value{dummy}}
\renewcommand{\theequation}{\thesection.\arabic{equation}}}
\def\d{{\partial}}

\def\ta{\hat a}
\def\tQ{\hat Q}
\def\tF{\hat F}

\begin{document}
\begin{flushright}
SISSA-ISAS 144/95/EP\\
IFT--P057/95\\
%hep-th/9506124
\end{flushright}
\vskip0.5cm
\centerline{\LARGE\bf Toda lattice realization of integrable hierarchies}
\vskip1cm
\centerline{\large  L.Bonora}
\centerline{International School for Advanced Studies (SISSA/ISAS)}
\centerline{Via Beirut 2, 34014 Trieste, Italy, and}
\centerline{INFN, Sezione di Trieste.  }
\vskip1cm
\centerline{\large C.P.Constantinidis}
\centerline{Instituto de F\'isica Te\'orica -- UNESP}
\centerline{Rua Pamplona 145, 01405 ${\rm S\tilde ao}$ Paulo, Brasil}
\vskip1cm
\centerline{\large E.Vinteler}
\centerline{International School for Advanced Studies (SISSA/ISAS)}
\centerline{Via Beirut 2, 34014 Trieste, Italy}
\vskip5cm
\abstract{We present a new realization of scalar integrable 
hierarchies in terms of the Toda lattice hierarchy. In other words, we show 
on a large number of examples that an integrable hierarchy, defined 
by a pseudodifferential Lax operator, can be embedded in the 
Toda lattice hierarchy. Such a realization
in terms the Toda lattice hierarchy seems to be as general as the 
Drinfeld--Sokolov realization.}
\vfill\eject

\section{Introduction}
\setcounter{equation}{0}
\setcounter{subsection}{0}

Scalar integrable hierarchies can be introduced in terms of
(pseudo)differential operators by means of a formalism first 
introduced by Gelfand and Dickey (see \cite{Dickey}). This is the 
most `disembodied' 
form in which such hierarchies can appear, and it can be taken 
as a reference form. One can then consider realizations of 
these hierarchies in physical systems.
A comprehensive realization is the one studied by 
Drinfeld and Sokolov in terms of linear systems defined on 
Lie algebras, \cite{DS}; let us refer
to it as the Drinfeld--Sokolov realization (DSR).
In this letter we present a new general realization of 
integrable hierarchies in terms of the Toda lattice hierarchy (TLH). 
We call it Toda lattice realization (TLR), and it looks as general 
as the DSR. While the DSR is contiguous to (reduced) WZNW models 
and Toda field theories in 2D, 
the TLR is inspired by matrix models, see \cite{BX1},\cite{BCX}.

The letter is organized as follows. In section 2 we introduce the TLR.
We do not give a general proof of it, but in section 3 we verify it 
on a large number of examples among KP, n--KdV and other classes 
of hierarchies. Section 5 is devoted to some comments.

\section{The Toda lattice realization of integrable hierarchies.}
\setcounter{equation}{0}
\setcounter{subsection}{0}

In the Gelfand--Dickey (GD) formalism an integrable hierarchy can 
be entirely specified in terms of the Lax operator
\a
L = \d^N + Na_1 \d^{N-2}+ Na_2 \d^{N-3}+...+Na_{N-1}+ N a_N \d^{-1}+...
\label{Lax}
\b
where $\d = {\d\over{\d x}}$. The operator $L$ may be purely 
differential, in which case $a_k=0$ for $ k\geq N$, and we get 
the $N$-KdV hierarchy. The fields $a_k$ 
may be either elementary or composite of more elementary fields,
as in the case of the $(N,M)$--KdV hierarchies studied in 
\cite{BX2},\cite{BX3},\cite{BLX}.
If the hierarchy is integrable, the flows are given by 
\a
\frac {\d L} {\d t_k} = [ (L^{k/N})_+, L] \label{GDflows}
\b
where the subscript + denotes the differential part of a 
pseudodifferential operator, $t_1$ is identified with $x$ and 
$k$ spans a specific subset of the positive integers.

The Toda lattice hierarchy is defined in terms of a semi--infinite 
Jacobi matrix $\hat Q$ \footnote{In this paper we limit ourselves 
to a simple version
of the TLH, in which only one matrix $\hat Q$ and one 
set of parameters intervene, 
instead of two or more \cite{UT},\cite{BX1}}. We parametrize it as 
follows 
\a
\hat Q=\sum_{j=0}^\infty \Big( E_{j,j+1} + \sum_{l=0}^{\infty} 
\hat a_l(j)E_{j+l,j}\Big), 
\qquad (E_{j,m})_{k,l}= \delta_{j,k}\delta_{m,l}\label{jacobi}
\b
and consider $\ta_l$ as fields defined on a lattice. The flows 
are given by
\a
\frac {\d \tQ} {\d t_k} = [(\tQ^k)_+ , \tQ] , 
\qquad k=1,2,...\label{Tflows}
\b
where the subscript + denotes the upper triangular part of a matrix, 
including the main diagonal. (\ref{Tflows}) represents a hierarchy of 
differential--difference equations for the fields $a_l$. In particular 
the first flows are 
\a
\ta_l(j)' = \ta_{l+1}(j+1)- \ta_{l+1}(j) + \ta_l(j)\Big(\ta_0(j)-
\ta_0(j-l)\Big)\label{ff}
\b
where we have adopted the notation $\frac {\d}{\d t_1}f \equiv f'\equiv \d f$, 
for any function $f$. The parameter $t_k$ of the TLH will be identified 
later on with the 
corresponding parameter $t_k$ in (\ref{GDflows}) whenever the latter exists; 
therefore, in particular, $t_1$ will be identified with $x$.

Next, integrability permits us to introduce the function $\tF(n,t)$ 
(the free energy in matrix models) via 
\a
\frac{\d^2}{\d t_{k}\d t_{l}} \tF(n,t) 
= \Tr\Big([\tQ^k_+, \tQ^l]\Big) 
\label{F}
\b 
where $\Tr (X)$ denotes the finite trace $\sum_{j=0}^{n-1}X_{j,j}$. In 
particular (\ref{F}) leads to
\a
\frac{\d^2}{\d t_{1}^2} \tF(n,t)=\ta_1(n)\label{a1}
\b 
It is clear that by means of (\ref{Tflows}) we can compute the derivatives
of any order of $\tF$ in terms of the entries of $\tQ$. In general we will
denote by $\tF_{k_1,...,k_s}$ the derivative of $\tF$ with respect to 
$t_{k_1},...,t_{k_s}$.

Next we introduce the operator $D_0$, defined by its action on any discrete 
function $f(n)$ 
\a
(D_0 f)(n) = f(n+1)\0
\b
For later use we remark that, if $f_0=0$, the operation 
$\Tr$ is the inverse of the operation $D_0-1$.
We will also use the notation $e^{\d_0}$ instead of $D_0$, 
with the following difference: $D_0$ is meant to be applied to the nearest 
right neighbour, while $e^{\d_0}$ acts on whatever is on its right.
Now we can equivalently represent the matrix $\tQ$ by the following operator
\a
\tQ(j) = e^{\d_0} + \sum_{l=0}^\infty \ta_l(j) e^{-l\d_0}\label{Qj}
\b
The contact between (\ref{Qj}) and (\ref{jacobi}) is made by acting with the 
former on a discrete function $\xi(j)$; then $\tQ(j)\xi(j)$ 
is the same as 
the $j$--th component of $\tQ \xi$, where $\xi$ is a column vector with 
components
$\xi(0),\xi(1),...$. We will generally drop the dependence on $j$ 
in (\ref{Qj}) and merge the two symbols.  

After this short introduction to the GD formalism and the Toda lattice 
hierarchy, let us come to the presentation of the TLR of the integrable 
hierarchy defined
by the Lax operator (\ref{Lax}), i.e. to the problem of embedding the latter
into the TLH. The prescription consists of several steps.

{\it Step 1}. In $\tQ$ we set $\ta_0=0$ and replace the first flows 
(\ref{ff}) 
with
\a
D_0\ta_1 = \ta_1,\qquad D_0 \ta_i= \ta_i + \ta_{i-1}',
\qquad i=2,3,...\label{ff'}
\b

{\it Step 2}. We compute
\a
\frac {\d \ta_1}{\d t_k} = \d \Tr\Big([\tQ_+, \tQ^k]\Big)\equiv\d \tF_{1,k}
\label{dta1}
\b
The right hand side will be a polynomial of the fields $\ta_k$ 
to which monomials of $D_0$ and $D_0^{-1}$ are applied. Next we substitute
the first flows (\ref{ff'}) to eliminate the presence of $D_0$. 
Examples:
\a
&&\tF_{1,1} = \ta_1, \0\\
&&\tF_{1,2} = (D_0+1)\ta_2 = 2\ta_2 + \ta_1',\label{A1k}\\
&&\tF_{1,3} = (D_0^2 + D_0 +1) \ta_3 + D_0\ta_1\ta_1+
 \ta_1\ta_1 +\ta_1D_0^{-1}\ta_1= 3\ta_3 +3\ta_2' +\ta_1'' +3\ta_1^2\0
\b
and so on.

Next we recall that 
\a
\tF_{1,k} = \frac{\d^2}{\d t_{k}\d t_{1}} \tF\0
\b
Using this and (\ref{A1k}), we can recursively write all the derivatives
of $\ta_l$ with respect to the couplings $t_k$ (and in particular the flows)
in terms of derivatives
of $\tF$, which, in turn, can be expressed as functions of the entries 
of $\tQ$. Example:
\a
\frac {\d }{\d t_k}\ta_2 = {1\over 2} \d \Tr \Big([\tQ^2_+, \tQ^k]\Big)
-{1\over 2} \frac{\d}{\d t_k} \ta_1'\0
\b
In general we will need all $\tF_{k_1,...,k_n}$. Here are some general 
formulas. Let us introduce the symbols $\ak_j$ as follows
 \a
 \tQ^k=\edk+ka_1\edkd+\ak_2\edkt+\ak_3\edkp+\ak_4\edkc+\ldots
 \b
The explicit form of the first few is:
 \a
 \ak_2&=&\left(\ba{c}k\\2\ea\right) a_1'+k\ta_2\label{Akj}\\
 \ak_3&=&\left(\ba{c}k\\3\ea\right) a_1''+\left(\ba{c}k\\2\ea\right)\ta_2'
+k\ta_3+\left(\ba{c}k\\2\ea\right) a^2_1\0\\
 \ak_4&=&\left(\ba{c}k\\4\ea\right) a_1'''+\left(\ba{c}k\\3\ea\right)\ta_2''+
\left(\ba{c}k\\2\ea\right)\ta_3'+
 k\ta_4+(3\left(\ba{c}k\\3\ea\right)-\left(\ba{c}k\\2\ea\right))a_1a_1'+
2\left(\ba{c}k\\2\ea\right) a_1\ta_2\label{Akk}\0\\
 \ak_5&=&\left(\ba{c}k\\5\ea\right) a_1^{(4)}+
\left(\ba{c}k\\4\ea\right)\ta_2'''+
\left(\ba{c}k\\3\ea\right)\ta_3''+
 \left(\ba{c}k\\2\ea\right) \ta_4'+k\ta_5+
\left(\ba{c}k\\2\ea\right) \ta_2\ta_2+
2\left(\ba{c}k\\2\ea\right) a_1\ta_3\0\\
&&+\left(\ba{c}k\\3\ea\right) a_1^3+
 (3\left(\ba{c}k\\4\ea\right)-\left(\ba{c}k\\3\ea\right))a_1'a_1'+
 (4\left(\ba{c}k\\4\ea\right)-2\left(\ba{c}k\\3\ea\right)+
\left(\ba{c}k\\2\ea\right))a_1a_1''+\0\\
&& (3\left(\ba{c}k\\3\ea\right)-\left(\ba{c}k\\2\ea\right)) a_1\ta_2'+
  (3\left(\ba{c}k\\3\ea\right)-2\left(\ba{c}k\\2\ea\right))\ta_2 a_1'\0
\b
and so on. In terms of these coefficients we can compute all the derivatives
of $\tF$. For example
\a
\tF_{1,k} &=& \ak_k\0\\
\tF_{2,k}&=&(D_0+1)\ak_{k+1}\0\\
 \tF_{3,k}&=&(D_0^2+D_0+1)\ak_{k+2}+3a_1\ak_k\label{Fkk}\\
\tF_{4,k}&=&(D_0^3+D_0^2+D_0+1)\ak_{k+3}+
4a_1(D_0+1)\ak_{k+1}+a_2^{(4)}\ak_{k}\0
\b
This procedure allows us to compute all the derivatives of the fields $\ta_l$ 
in terms of the same fields and their derivatives with respect to 
$x\equiv t_1$ -- therefore, in particular, the flows.

So far all our moves have been completely general (except for setting 
$\ta_0=0$,
but see the comment at the end of this section). The next step is instead
a `gauge choice', that is we make a particular choice for the matrix $\tQ$.
The word `gauge' is not merely colorful. In fact gauge transformations play
here a role analogous to gauge transformations in \cite{DS}. The relevant
gauge transformations in the present case are defined by $\tQ$ $\to$ 
$G_-\tQ G_-^{-1}$, where $G_-$ is a strictly lower triangular semi--infinite
matrix.

{\it Step 3}. We fix the gauge by imposing the condition
\a
\tQ^N= e^{N\d_0} + \sum_{l=1}a_l e^{(N-1-l)\d_0}\label{gf}
\b
where the $a_l$ are the same as in eq.(\ref{Lax}).
The matrix $\tQ$ that satisfies such condition will be referred to as $\bar Q$. 
It is clear that $\bar Q^N$ exactly mimics the Lax operator $L$. 
The condition (\ref{gf}) recursively determines   
$\ta_k$ in terms of the fields $a_l$ that appear in $L$.
\a
\ta_k = \bar a_k \equiv P_k(a_l)\0
\b
where $P_k$ are differential polynomials of $a_l$.
In particular we always have $\ta_1= \bar a_1\equiv a_1$.

{\it Step 4}. Then we evaluate both sides of the flows found in {\it Step 2} 
at $ \ta_k = \bar a_k $. The order here is crucial. The gauge fixing of the
flows must be the last operation.

Now we claim:

{\bf Claim}. {\it The flows obtained in this way coincide with the flows 
(\ref{GDflows}) for corresponding couplings.}

We will substantiate this claim with a large number of examples in the next 
section. 

It is perhaps useful to summarize our method: {\it start from the TLH flows, 
use the first flows (\ref{ff'}) and impose the relevant gauge fixing; the 
resulting flows are the desired differential integrable flows.}

We would like to end this section with a remark concerning the restriction
$\ta_0=0$ we imposed at the very beginning. 
This can be avoided at the price of
working with very encumbering formulas. One can keep 
$\ta_0\neq 0$ provided one
uses the first flows (\ref{ff}) instead of (\ref{ff'}) in {\it Step 1}.
In this way it is possible, in general, to eliminate $D_0$ in the flows 
only when it acts over $\ta_l$, $l\neq 0$ (see the last section for an 
additional comment on this point). We obtain in this way the same equations
as above with the addition of terms involving $\ta_0$. 
We can suppress all these
additional terms at the end ({\it Step 5}) by imposing $\ta_0=0$ as part of 
the gauge choice. The final result is of course the same as before. This 
justifies our having imposed $\ta_0=0$ from the very beginning.

\section{Examples.} 
\setcounter{equation}{0}
\setcounter{subsection}{0}

In this section we present a large number of examples in support of 
the claim of the previous section. Of course for obvious reasons of space
we can explicitly exhibit a few cases only, and for each case only a few flows
among those we have checked.

\subsubsection*{The KP hierarchy}

The KP case corresponds to $n=1$ in (\ref{Lax}). Therefore there is no gauge
fixing: $\ta_l=a_l$. The flows obtained with our method are simply those 
in {\it Step 3}. Examples:
\a
&&\frac{\d \ta_1} {\d t_2} = \Big(2\ta_2+\ta_1'\Big)', \qquad\qquad
\frac{\d \ta_1} {\d t_3} = \Big(3\ta_3 +3\ta_2' +\ta_1'' 
+3\ta_1^2\Big)'\0\\
&&\frac{\d \ta_2} {\d t_2}= \Big(2\ta_3 +\ta_2' + \ta_1^2\Big)',\qquad
\frac{\d \ta_2} {\d t_3}= \Big(3\ta_4 +3\ta_3' +\ta_2'' +6 \ta_1
\ta_2\Big)'\0
\b
and so on. Setting $\ta_l=a_l$, these are exactly the KP flows.

\subsubsection*{The $N$-KdV hierarchy case}

In \cite{BCX} we have explicitly shown that our claim is true for the 3--KdV
hierarchy. In this section we generalize that result. To start with we 
pick a generic $N$. The relevant differential operator is 
\a
 L=D^N+Na_1D^{N-2}+Na_2D^{N-3}+\ldots +Na_{N-1}\label{nkdv}
\b
We also write
\a
L^{k/N}=D^k+ka_1D^{k-2}+b_2^{[k]}D^{k-3}+\ldots +b_{j-1}^{[k]}D^{k-j}
+...\label{Lk}
\b
The coefficients $b_j^{[k]}$ are differential polynomials in $a_l$, 
$l=1,...,a_{N-2}$.

Working out the commutator in relation (\ref{GDflows}), we can
write down the general formula for arbitrary flow $t_m$:
\a
{\d a_{j-1}\over \d t_m}&=&
\sum_{k=0}^{m-1}\left(\ba{c}m\\k\ea\right)a_{j+k-1}^{(m-k)}-
\sum_{k=0}^{m-2}{1\over N}\left(\ba{c}N\\j+k\ea\right)
(b_{m-k-1}^{(m)})^{(j+k)}+\label{nflows}\\
&+&\sum_{k=0}^{m-3}(\left(\ba{c}m-2\\k\ea\right)
b_{k-1}^{(m)}a_{j-k-1}^{(m-2-k)}-
\sum_{l=0}^{m-2}\left(\ba{c}N-j-k\\l-k\ea\right)
a_{k-1}(b_{m-l-1}^{(m)})^{(l-k)})\0\\
&-&\sum_{k=2}^{j-1}\sum_{l=0}^{m-2}\left(\ba{c}N-k\\j-k+l\ea\right) a_{k-1}
(b_{m-l-1}^{(m)})^{(j-k+l)}\0
\b

Now let us pass to the TLR of this hierarchy. We recall eqs.(\ref{Akj})
and (\ref{Fkk}). We fix the gauge by imposing $\an_j=Na_j$. We solve
the equations for $\ta_j$ in terms of $a_j$ and obtain $\bar a_j$. Next we
insert back the result in the formulas of the coefficients $\ak_j$ so that
they become functions of $a_j$. We call the result $\bak_j$.
Examples:
 \a
 \bak_2/k&=&a_2-{N-k\over 2}a_1'\0\\
 \bak_3/k&=&a_3-{N-k\over 2}a_2'+{(N-2k+3)(N-k)\over 12}a_1''- 
 {N-k\over 2}a_1^2\\
\bak_4/k&=&a_4-{N-k\over 2}a_3'+{(N-2k+3)(N-k)\over 12}a_2''-
{(N-k+2)(k-2)(N-k)\over 24}a_1'''\0\\
&&- (N-k)a_1 a_2+{(N-k+2)(N-k)\over 2}a_1 a_1'\0
 \b

Then, using our recipe, we obtain
\a
 \d^{-1}{\d a_1\over\d t_k}&=&\frac {\d^2 F}{\d t_1\d t_k}|_{\hat a=\bar a}
 =\bak_k\0\\
 \d^{-1}{\d a_2\over\d t_k}&=&\Big({1\over 2}
  \frac {\d^2 F}{\d t_2\d t_k}\Big)|_{\hat a=\bar a}
 +{N-2\over 2}{\d a_1\over \d t_k}=
     \bak_{k+1}+ {N-1\over 2}(\bak_k)'\0\\
 \d^{-1}{\d a_3\over\d t_k}&=&\Big({1\over 3} 
 \frac {\d^2 F}{\d t_3\d t_k}\Big)|_{\hat a=\bar a}+
 {N-3\over 2}{\d a_2\over \d t_k}-{(N-3)^2\over 12}{\d a_1'\over \d t_k}
   +(N-3)\d^{-1}(a_1{\d a_1\over \d t_k})\0\\
   &=& \bak_{k+2}+ {N-1\over 2}(\bak_{k+1})'+{(N-1)(N-2)\over 6}(\bak_k)''+
   a_1 \bak_k+(N-3)\d^{-1}(a_1(\bak_k)')\0\\
\d^{-1}{\d a_4\over\d t_k}&=&\Big({1\over 4} 
\frac {\d^2 F}{\d t_4\d t_k}+
{N-4\over 2}{\d a_3\over \d t_k}\Big)|_{\hat a=\bar a}
-{(N-4)(N-5)\over 12}{\d a_2'\over \d t_k}
+(N-4)\d^{-1}{\d( a_1a_2)\over\d t_k}\0\\
&-& {(N-4)(N-2)\over 2}({1\over 6}{\d a_1''\over \d t_k}
+a_1{\d a_1\over \d t_k})=
 \bak_{k+3}+ {N-1\over 2}(\bak_{k+2})'\0\\
&+&{(3N-11)(N-1)(N-2)\over 24}(\bak_{k})'''-
{(N-2)(N-4)\over 2}a_1(\bak_k)'+\0\\
&+&{(5N-16)(N-1)\over 12}(\bak_{k+1})''+ (N-4)\d^{-1}(a_1 (\bak_{k+1})'+
{N-1\over 2}(a_1(\bak_{k})''+
a_2(\ak_k)')\0
\label{toda234}
\b
and so on, where ${\hat a=\bar a}$ denotes gauge fixing. 

We give a few concrete examples of the second and third flows:
\a
\d^{-1}{\d a_1\over \d t_2}&=&2a_2-(N-2)a_1'\0\\
\d^{-1}{\d a_2\over \d t_2}&=&2a_3+a_2'-{(N-1)(N-2)\over 3}a_1''-
(N-2)a_1^2\0\\
{\d a_3\over \d t_2}&=&2a_4'+a_3''-{(N-1)(N-2)(N-3)\over 12}a_1^{(4)}
-(N-2)(N-3)a_1a_1''-2(N-3)a_2a_1'\0
\b
\a
\d^{-1}{\d a_1\over \d t_3}&=&3a_3-{3\over 2}(N-3)a_2'+
{(N-3)^2\over 4}a_1''-{3\over 2}(N-3)a_1^2\0\\
\d^{-1}{\d a_2\over \d t_3}&=&3a_4+3a_3'-{N(N-3)\over 2}a_2''+
{(N-1)(N-2)(N-3)\over 8}a_1'''-3(N-3)a_1a_2\0\\
{\d a_3\over \d t_3}&=&3a_5'+3a_4''+a_3'''-
{3\over N}\left(\ba{c}N\\4\ea\right)a_2^{(4)}+
{3(3N-7)\over 10N}\left(\ba{c}N\\4\ea\right)a_1^{(5)}+
3a_1a_3'-3(N-4)a_1'a_3\0\\
&&-3(N-3)a_2a_2'-{3\over 2}(N-2)(N-3)a_1a_2''+
{3\over 2}(N-3)a_2a_1''+
{6\over N}\left(\ba{c}N\\4\ea\right)a_1a_1'''\0
\b
These are flows pertinent to the $N$--KdV hierarchy with $N>3$.
In general the formulas of the $N$--KdV hierarchy and the corresponding
formulas obtained with our method coincide since
\a
b_k^{[m]}=\bar A_k^{[m]}\0
\b
We have checked these identities case by case up to the 5-KdV and 
for $m\leq 5$. For the dispersionless case we have verified the 
correspondence up to the 8--KdV flows.

\subsubsection*{The DS hierarchies}
 
Drinfeld and Sokolov, \cite{DS}, introduced a large set of 
generalized KdV systems in terms of 
the pair $(G,c_m)$, where $G$ is a classical Kac-Moody algebra and
$c_m$ is a vertex of the Dynkin diagram of $G$. From each choice
of  the pair $(G,c_m)$ they were able
to construct a pseudo--differential operator $L$ which give rise to a 
hierarchy of integrable equations.
We have studied all the examples corresponding to the operator 
$L$ of orders 3,4,5 and found a complete agreement with our method. 
For simplicity here we present a few examples of order 4 and 5, corresponding
to the cases with a pseudodifferential Lax operator. The 
cases with a differential Lax operator are restriction of the 4-- and 5--KdV
hierarchies, and will be omitted.

In each case we give the
explicit form of the (pseudo--)differential operator $L$, the 
gauge--fixed matrix $\bar Q$ and the first significant flows:

\noindent{\bf Order 4.}
\vskip 0.2 cm
\noindent {\bf  Case} $B_2^{(1)}$:
\a
c_0,c_1~~:L&=&D^4+2u_1D^2+u_1'D+2(u_0+u_1'')-D^{-1}(u_0+u_1'')'\\
\tQ &=&\ed+{v_1\over 4}\em-{v_1'\over 4}\emd+({1\over 4}v_0+{v_1''\over 8}-
{3\over 32}v_1^2)\emt+\0\\
&&+({3\over 8}v_1v_1'-{1\over 2}v_0')\emp+\ldots\0\\
c_2~~:L&=&D^4+2u_1D^2+u_1'D+u_0^2-u_0D^{-1}u_0'\0
\b
For $c_2$ we have, up to the order $\emp$, the same expression for $\tQ$ 
with $v_0=u_0^2$. The first non--trivial flows are:
\a
{\d v_1\over \d t_3}&=&-{1\over 2}v_1'''-{3\over 4}v_1v_1'+3v_0'\0\\
{\d v_0\over \d t_3}&=&v_0'''+{3\over 4}v_1v_0'
\b
where for $c_0,c_1:v_1=2u_1,v_0=2(u_0+u_1'')$  and for 
$c_2:~v_1=2u_1,v_0=u_0^2$.

\noindent{\bf Case} $D_3^{(1)}$:
\a
c_0,c_1:~~L&=&D^4+2u_2D^2+u_2'D+2u_2''+2u_1-D^{-1}(u_1'+u_2'')'+
        (D^{-1}u_0)^2\label{D31}\\
\tQ&=&\ed+{u_2\over 2}\em-{1\over 2}u_2'\emd+
({3\over 4}u_2''-{3\over 8}{u_2^2}-{1\over 2}u_1)\emt+\0\\
&&+({3\over 2}u_2u_2'-u_1'-u_2''')\emp+\ldots\0\\
c_2,c_3:~~L&=&D^4+2u_2D^2+3u_2'D+(2u_1+3u_2'')+(u_1'+u_2''')D^{-1}+
u_0D^{-1}u_0D^{-1}\0\\
\tQ&=&\ed+{u_2\over 2}\em+
({1\over 4}u_2''-{3\over 8}u_2^2+{1\over 2}u_1)\emt+\0\\
&&+({3\over 4}u_2u_2'-{1\over 2}u_1'-{1\over 4}u_2''')\emp+\ldots\0
\b
The first non--trivial equations for $c_0,c_1$ are:
\a
{\d u_1\over \d t_3}&=&3u_0u_0'-{3\over 2}u_2^{(5)}-2u_1'''+{3\over 2}u_1'u_2
+3u_2u_2'''+{9 \over 2}u_2'u_2''\label{fD31}\\
{\d u_2\over \d t_3}&=&{5\over 2}u_2'''-{3\over 2}u_2u_2'+3u_1'\0
\b
and for $c_2,c_3$ are
\a
{\d u_1\over \d t_3}&=&3u_0u_0'-{3\over 2}u_2^{(5)}-2u_1'''-
{3\over 2}u_1'u_2+{9\over 2}u_2'u_2''+3u_2u_2'''\label{fD31'}\\
{\d u_2\over \d t_3}&=&{5\over 2}u_2'''-{3\over 2}u_2u_2'+3u_1'\0
\b

\vskip.2cm
\noindent {\bf Order 5.}
\vskip.2cm
 
\noindent {\bf Case} $A_5^{(2)}$:
\a
c_0,c_1~~:&&L= D^5+2u_2D^3+2u_2'D^2+(2u_1+4u_2'')D+D^{-1}(2u_0+u_1''+u_2''')\\
&&\tQ=\ed+{2\over 5}u_2\em-{2\over 5}u_2'\emd+
{2\over 5}(u_1+2u_2''-{4\over 5}u_2^2)\emt+\ldots\0\\
c_2~~:&&L=D^5+2(v_0+u_1)D^3+(6v_0'+u_1')D^2+(6v_0''+u_0^2+4v_0u_1)D+\0\\
&&+(2v_0'''-u_0u_0'+4u_1v_0'+2v_0u_1')+u_0D^{-1}(u_0''+2u_0v_0)\0\\
&&\tQ=\ed+{2\over 5}(u_1+v_0)\em+{1\over 5}(2v_0'-3u_1')\emd+\0\\
&&+{1\over 5}(2u_1''-2v_0''+{4\over 5}v_0u_1+u_0^2-
{8\over 5}u_1^2-{8\over 5} v_0^2)\emt+\ldots\0\\
c_3~~:&&L=D^5+2u_2D^3+3u_2'D^2+(2u_1+3u_2'')D+u_1'+u_2'''+u_0D^{-1}u_0\0\\
&&\tQ=\ed+{2\over 5}u_2\em-{1\over 5}u_2'\emd+{1\over 5}
(2u_1+u_2''-{8\over 5}u_2^2)\emt+\ldots\0
\b
The equations are for $c_0,c_1$:
\a
{\d u_2\over \d t_3}&=&4u_2'''+3u_1'-{12\over 5}u_2u_2'\\
{\d u_1\over \d t_3}&=&-{7\over 2}u_1'''+6u_2u_2'''+{54\over 5}u_2'u_2''
+{6\over 5}(u_2u_1'-u_1u_2')-{{51}\over {10}}u_2^{(5)}+3 u_0'\0
\b
for $c_2$:
\a
5{\d v_0\over \d t_3}&=&-v_0'''+
{3\over 2}u_1'''-12v_0v_0'+6v_0'u_1+12v_0u_1'\\
5{\d u_1\over \d t_3}&=&6v_0'''-
4u_1'''+15u_0u_0'+6v_0u_1'+12v_0'u_1-12u_1u_1'\0
\b
and for $c_3$ are:
\a
{\d u_2\over \d t_3}&=&u_2'''+3u_1'-{12\over 5}u_2u_2'\\
{\d u_1\over \d t_3}&=&-2u_1'''+{27\over 5}u_2'u_2''+{12\over 5}u_2u_2'''
+{6\over 5}(u_2u_1'-u_1u_2')-{3\over 5}u_2^{(5)}+3u_0u_0'\0
\b

\subsubsection*{The (N,M)--KdV hierarchies}

The $(N,M)$--KdV hierarchies are defined by the pseudodifferential 
operator
\a
L=\d^{N}+N\sum_{l=1}^{N-1} a_l\d^{N-l-1}
 +N\sum_{l=1}^M a_{N+l-1}{1\over{\d-S_l}}{1\over{\d-S_{l-1}}}
  \ldots {1\over{\d-S_1}},\quad N\ge1,~M\ge0 \label{pdo}
\b
The case $(N,0)$ coincides with the $N$--KdV case. These hierarchies 
were studied in \cite{BX2},\cite{BX3},\cite{BLX},\cite{D}. In \cite{BLX}
it was shown that they can be embedded in the DS construction. Now we show
that this class of integrable hierarchies can be entirely embedded in the
TLH. Let us see, for example, the $(2,1)$ case. The Lax operator is 
\a
L=\d^2+2a_1+2a_2\frac{1}{\d-S}\0
\b
The gauge fixing gives
\a
\bar a_1= a_1,\qquad\bar a_2 = a_2 -{1\over 2} a_1',\qquad
&&\bar a_3 = - {1\over 2}a_2' +{1\over 4} a_1''-{1\over 2}a_1^2
+a_2S\0
\b
and so on. It leads, via our recipe, to the following flows
\a
&&\d^{-1} \frac{\d a_1} {\d t_2} = 2a_2,\qquad
\d^{-1} \frac{\d a_2} {\d t_2} = a_2' + 2a_2S,\qquad
\d^{-1} \frac{\d S} {\d t_2}= S^2 + 2a_1 -S,'\0\\
&&\d^{-1} \frac{\d a_1} {\d t_3}= {3\over 2} a_2' +{1\over 4}a_1''
+{3\over 2}a_1^2 +3 a_2S,\qquad
\d^{-1} \frac{\d a_2} {\d t_3}= a_2''+3a_1a_2+3a_2'S+3a_2S^2\0
\b
and so on. These are exactly the flows of the (2,1)--KdV hierarchy.

\section{Comments and conclusion}. 
\setcounter{equation}{0}
\setcounter{subsection}{0}

The examples we have considered in the previous section do not 
exhaust all possible integrable hierarchies (for an updating on this
subject see \cite{Grimm}). However they are very numerous and they
leave very little doubt that whatever scalar Lax operator (\ref{Lax}),
defining an integrable hierarchy, we may think of, it can be embedded
in the Toda lattice hierarchy in the way we showed above. 
Anyhow, thus far we have not found any counterexample.
Therefore our construction looks at
least as general as the DS realization. The fact that we are dealing
with semi--infinite matrices may suggest additional possibilities.

We also remark that the TLH, in its general formulation, may
encompass several $\tQ$ matrices (not only one, as in this paper).
Therefore there is room for `tensor products of integrable hierarchies
in interaction'.

We end the paper by recalling that in the case of the 
${(1,M)}$--KdV hierarchies
there is a variant to the realization of section 2. This was already
pointed out in section 6.2 of \cite{BCX} and, implicitly, in \cite{BX2}. 
If one does not set $\ta_0=0$ and replaces the first flows (\ref{ff}) 
in the Toda lattice flows, one gets exactly the $(1,M)$ hierarchies if
the gauge fixing simply consists of setting $\ta_l=0$ for $l>M$.
It was shown in \cite{BX2} 
that $(N,M)$--KdV hierarchies can then be extracted from the $(1,M)$ 
via a cascade Hamiltonian reduction. However 
it is not clear whether this method can be generalized to other hierarchies,
and, anyhow, it does not seem to be appropriate to call it a realization
of differential hierarchies, at least in the same sense this terminology 
has been used in this paper.

\vskip.5cm
{\bf Acknowledgements}. One of us (C.P.C.) would like to thank CNPq and FAPESP
for financial support.

\end{document}